\documentclass[twocolumn]{aastex63}
\hypersetup{linkcolor=red,citecolor=blue,filecolor=cyan,urlcolor=black}

\shorttitle{GLASS-JWST: Abell~2744 NIRCam photometric catalog}
\shortauthors{Paris et al.}

\begin{document}

\title{The GLASS-JWST Early Release Science Program. II. Stage I release of NIRCam imaging and catalogs in the Abell~2744 region.}

\correspondingauthor{Diego Paris}
\email{diego.paris@inaf.it}

\author[0000-0002-7409-8114]{Diego Paris}
\affiliation{INAF Osservatorio Astronomico di Roma, Via Frascati 33, 00078 Monteporzio Catone, Rome, Italy}

\author[0000-0001-6870-8900]{Emiliano Merlin}
\affiliation{INAF Osservatorio Astronomico di Roma, Via Frascati 33, 00078 Monteporzio Catone, Rome, Italy}

\author[0000-0003-3820-2823]{Adriano Fontana}
\affiliation{INAF Osservatorio Astronomico di Roma, Via Frascati 33, 00078 Monteporzio Catone, Rome, Italy}

\author[0000-0002-2667-5482]{Andrea Bonchi}
\affiliation{Space Science Data Center, Italian Space Agency, via del Politecnico, 00133, Roma, Italy}
\affiliation{INAF Osservatorio Astronomico di Roma, Via Frascati 33, 00078 Monteporzio Catone, Rome, Italy}

\author[0000-0003-2680-005X]{Gabriel Brammer}
\affiliation{Cosmic Dawn Center (DAWN), Denmark}
\affiliation{Niels Bohr Institute, University of Copenhagen, Jagtvej 128, DK-2200 Copenhagen N, Denmark}

\author[0000-0001-6464-3257]{Matteo Correnti}
\affiliation{Space Science Data Center, Italian Space Agency, via del Politecnico, 00133, Roma, Italy}
\affiliation{INAF Osservatorio Astronomico di Roma, Via Frascati 33, 00078 Monteporzio Catone, Rome, Italy}

\author[0000-0002-8460-0390]{Tommaso Treu}
\affiliation{Department of Physics and Astronomy, University of California, Los Angeles, 430 Portola Plaza, Los Angeles, CA 90095, USA}

\author[0000-0003-4109-304X]{Kristan Boyett}
\affiliation{School of Physics, University of Melbourne, Parkville 3010, VIC, Australia}
\affiliation{ARC Centre of Excellence for All Sky Astrophysics in 3 Dimensions (ASTRO 3D), Australia}

\author[0000-0003-2536-1614]{Antonello Calabr\`o}
\affiliation{INAF Osservatorio Astronomico di Roma, Via Frascati 33, 00078 Monteporzio Catone, Rome, Italy}

\author[0000-0001-9875-8263]{Marco Castellano}
\affiliation{INAF Osservatorio Astronomico di Roma, Via Frascati 33, 00078 Monteporzio Catone, Rome, Italy}

\author[0000-0003-1060-0723]{Wenlei Chen}
\affil{Minnesota Institute for Astrophysics, University of Minnesota, 116 Church Street SE, Minneapolis, MN 55455, USA}

\author[0000-0002-8434-880X]{Lilan Yang}
\affiliation{Kavli Institute for the Physics and Mathematics of the Universe, The University of Tokyo, Kashiwa, Japan 277-8583}

\author[0000-0002-3254-9044]{K. Glazebrook}
\affiliation{Centre for Astrophysics and Supercomputing, Swinburne University of Technology, PO Box 218, Hawthorn, VIC 3122, Australia}

\author[0000-0003-3142-997X]{Patrick Kelly}
\affil{Minnesota Institute for Astrophysics, University of Minnesota, 116 Church Street SE, Minneapolis, MN 55455, USA}

\author[0000-0002-6610-2048]{Anton M. Koekemoer}
\affiliation{Space Telescope Science Institute, 3700 San Martin Dr., Baltimore, MD 21218, USA}

\author[0000-0003-4570-3159]{Nicha Leethochawalit}
\affiliation{National Astronomical Research Institute of Thailand (NARIT), Mae Rim, Chiang Mai, 50180, Thailand}

\author[0000-0002-9572-7813]{Sara Mascia}
\affiliation{INAF Osservatorio Astronomico di Roma, Via Frascati 33, 00078 Monteporzio Catone, Rome, Italy}

\author[0000-0002-3407-1785]{Charlotte Mason}
\affiliation{Cosmic Dawn Center (DAWN), Denmark}
\affiliation{Niels Bohr Institute, University of Copenhagen, Jagtvej 128, DK-2200 Copenhagen N, Denmark}

\author[0000-0002-8512-1404]{Takahiro Morishita}
\affiliation{IPAC, California Institute of Technology, MC 314-6, 1200 E. California Boulevard, Pasadena, CA 91125, USA}

\author[0000-0001-6342-9662]{Mario Nonino}
\affiliation{(INAF - Osservatorio Astronomico di Trieste, Via Tiepolo 11, I-34131 Trieste, Italy)}

\author[0000-0001-8940-6768 ]{Laura Pentericci}
\affiliation{INAF Osservatorio Astronomico di Roma, Via Frascati 33, 00078 Monteporzio Catone, Rome, Italy}

\author[0000-0003-4067-9196]{Gianluca Polenta}
\affiliation{Space Science Data Center, Italian Space Agency, via del Politecnico, 00133, Roma, Italy}

\author[0000-0002-4140-1367]{Guido Roberts-Borsani}
\affiliation{Department of Physics and Astronomy, University of California, Los Angeles, 430 Portola Plaza, Los Angeles, CA 90095, USA}

\author[0000-0002-9334-8705]{Paola Santini}
\affiliation{INAF Osservatorio Astronomico di Roma, Via Frascati 33, 00078 Monteporzio Catone, Rome, Italy}

\author[0000-0001-9391-305X]{Michele Trenti}
\affiliation{School of Physics, University of Melbourne, Parkville 3010, VIC, Australia}
\affiliation{ARC Centre of Excellence for All Sky Astrophysics in 3 Dimensions (ASTRO 3D), Australia}

\author[0000-0002-5057-135X]{Eros Vanzella}
\affiliation{INAF -- OAS, Osservatorio di Astrofisica e Scienza dello Spazio di Bologna, via Gobetti 93/3, I-40129 Bologna, Italy}

\author[0000-0003-0980-1499]{Benedetta Vulcani}
\affiliation{INAF Osservatorio Astronomico di Padova, vicolo dell'Osservatorio 5, 35122 Padova, Italy}

\author[0000-0001-8156-6281]{Rogier A. Windhorst} 
\affiliation{School of Earth and Space Exploration, Arizona State University,
Tempe, AZ 85287-1404, USA}

\author[0000-0003-2804-0648 ]{Themiya Nanayakkara}
\affiliation{Centre for Astrophysics and Supercomputing, Swinburne University of Technology, PO Box 218, Hawthorn, VIC 3122, Australia}

\author[0000-0002-9373-3865]{Xin Wang}
\affil{School of Astronomy and Space Science, University of Chinese Academy of Sciences (UCAS), Beijing 100049, China}
\affil{National Astronomical Observatories, Chinese Academy of Sciences, Beijing 100101, China}
\affil{Institute for Frontiers in Astronomy and Astrophysics, Beijing Normal University,  Beijing 102206, China}

\begin{abstract}
We present images and a multi--wavelength photometric catalog based on all of the $JWST$ NIRCam observations obtained to date in the region of the Abell~2744 galaxy cluster. These data come from three different programs, namely the GLASS-JWST Early Release Science Program, UNCOVER, and Director's Discretionary Time program 2756. The observed area in the NIRCam wide-band filters - covering the central and extended regions of the cluster, as well as new parallel fields - is 46.5 arcmin$^2$ in total.
All images in eight bands (F090W, F115W, F150W, F200W, F277W, F356W, F410M, F444W) have been reduced adopting the latest calibration and reference files available. Data reduction has been performed using an augmented version of the official $JWST$ pipeline, with improvements aimed at removing or mitigating defects in the raw images and improving the background subtraction and photometric accuracy. We obtain an F444W-detected multi--band catalog, including all NIRCam and available $HST$ data, adopting forced aperture photometry on PSF-matched images. The catalog is intended to enable early scientific investigations and is optimized for the study of faint galaxies; it contains 24389 sources, with a 5$\sigma$ limiting magnitude in the F444W band ranging from 28.5 AB to 30.5 AB, as a result of the varying exposure times of the surveys that observed the field. 
We publicly release the reduced NIRCam images, associated multi-wavelength catalog and code adopted for $1/f$ noise removal with the aim of aiding users to familiarize themselves with $JWST$ NIRCam data and identify suitable targets for follow-up observations.

\end{abstract}


\section{Introduction}
\label{sec:intro}

In just a few months of observations, $JWST$ has demonstrated its revolutionary scientific capabilities. Early observations have shown  that its performance is equal or better than expected, with image quality and overall efficiency that matches or surpasses pre-launch estimates \citep{Rigby2022}.  Publicly available datasets obtained by the Early Release Observations and Early Release Science programs have already enabled a large number of publications based on $JWST$ data, ranging from exoplanets to the distant Universe.

In particular, many works exploited the power of NIRCam to gather the first sizeable sample of candidates at $z\geq 10$ \citep[e.g.,][]{Castellano2022b,Donnan2023,Finkelstein2022b,morishita22a,Naidu2022b,Yan2022,RobertsBorsani2022d,Robertson2022b,Bouwens2022,Castellano2023}, demonstrating the power of $JWST$ in exploring the Universe during the re-ionization epoch.

In this paper, we present the full data set obtained with NIRCam in the region of the $z=0.308$ cluster Abell~2744 that will significantly expand the available area for deep extragalactic observations. The central region of the cluster allows an  insight into the distant Universe at depth and resolution superior to those of NIRCam in blank fields, with the lensing magnification assistance. The data analyzed here are obtained through three public programs: i) GLASS JWST-ERS-1324 \citep{Treu2022}, ii) UNCOVER JWST-GO-2561 \citep{Bezanson2022}, and iii) Director's Discretionary Time Program 2756, aimed at following up a Supernova discovered in GLASS-JWST NIRISS imaging. We have analyzed and combined the imaging data of all these programs and obtained a multi-wavelength catalog of the objects detected in the F444W band.

In order to facilitate the exploitation of these data, we release reduced images and associated catalog on our website and through the Mikulski Archives for Space Telescopes (MAST). This release fulfills and exceeds the requirements of the Stage I data release planned as part of the GLASS-JWST program. It is anticipated that a final (Stage II) release will follow in approximately a year, combining additional images scheduled in 2023, and taking advantage of future improvements in data processing and calibrations.

This paper is organized as follows. In Section~\ref{sec:data} we present the data-set and discuss the image processing pipeline. In Section~\ref{sec:catalog}, the methods applied for the detection of the sources and the photometric techniques used to compute the fluxes are presented. Finally in Section~\ref{sec:conclusions}, we summarize the results. Throughout the paper we adopt AB magnitudes~\citep{oke83}.

\section{Data reduction}
\label{sec:data}
\subsection{Data Set}
\label{sec:dataset}

\begin{figure*}
\center
 \includegraphics[width=\textwidth]{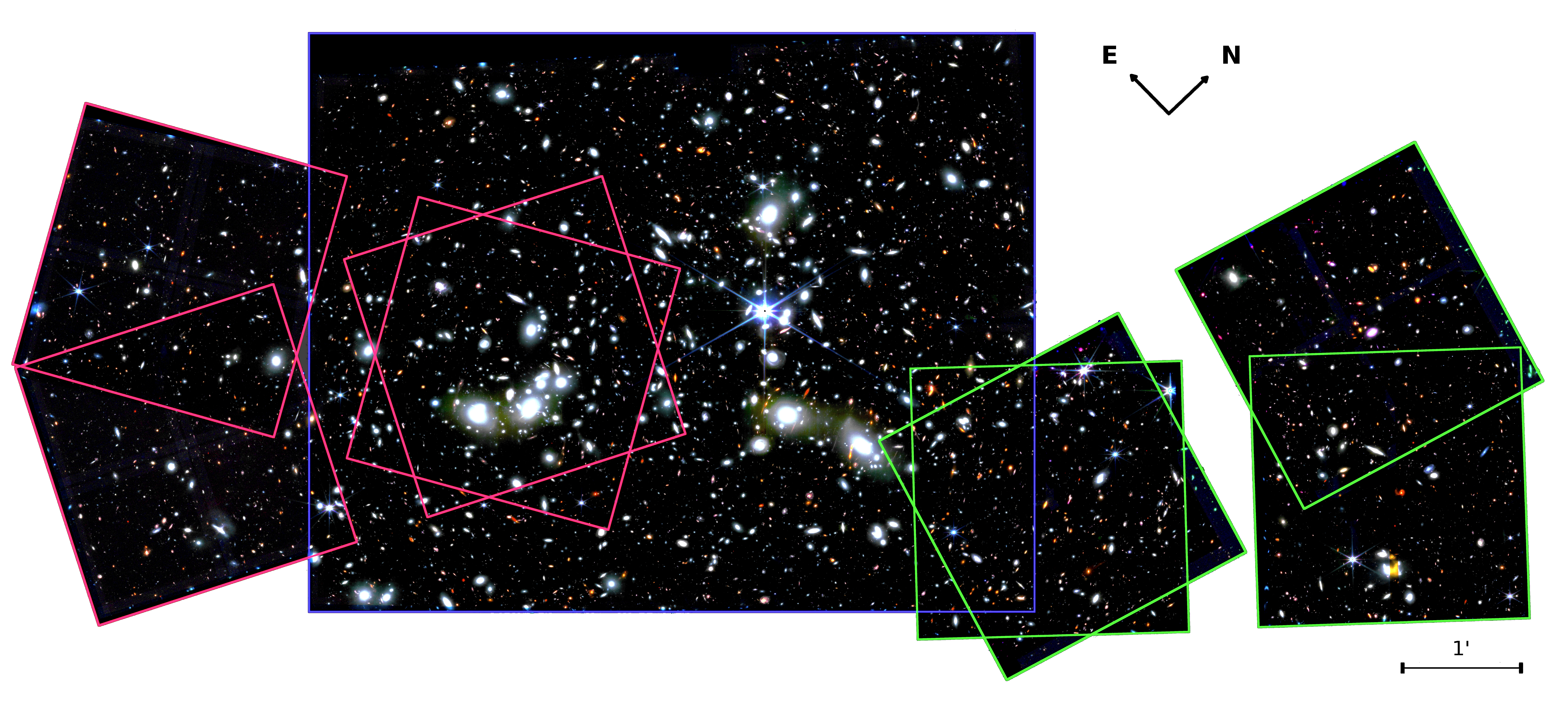}
\caption{Full view of the color composite RGB mosaic obtained combining the F090W+F115W+F150W as blue, F200W+F277W as green and F356W+F410M+F444W as red. Colored boxes show the position of the three different data sets used here: GLASS (green), UNCOVER (blue) and DDT (red). The entire image (including the empty space) is approximately $12.7\times5.9$ arcmin wide. }
 \label{fig:mosaic}
\end{figure*}

The NIRCam data analyzed in this paper are taken from three programs that targeted the $z=0.308$ cluster Abell~2744 (A2744 hereafter) and its surroundings. The first set of NIRCam images were taken as part of the GLASS-JWST survey \citep{Treu2022}, in parallel to primary NIRISS observations on June 28--29, 2022 and to NIRSpec observations on Nov. 10--11, 2022. We refer to these data sets as GLASS1 and GLASS2, or collectively as GLASS,  both of which consist of imaging in seven broad-band filters from F090W to F444W (see \citep{Treu2022}, hereafter T22, for details). We note that the final pointing is different from the scheduled one presented by T22 due to the adoption of an alternate position angle (PA) during the NIRSpec spectroscopic observations. As the primary spectroscopic target was the A2744 cluster, these parallel images are offset to the North-West. By virtue of the long exposure times, these images are the deepest presented here.

The second set of NIRCam observations considered here were taken as part of the UNCOVER program \citep{Bezanson2022}, which targets the center of the A2744 cluster and the immediate surroundings. These images are composed of four pointings and result in a relatively homogeneous depth, as discussed below. They were taken on November 2-4-7 and 15 and adopt the same filter set as GLASS-JWST, except for the addition of the F410M filter instead of F090W.

Finally, DDT program 2756 (PI W. Chen, DDT hereafter) also obtained NIRCam imaging data in the center of A2744 on October 20 and December 6, 2022 (UT). These two data sets are dubbed DDT1 and DDT2 hereafter.
The DDT set-up is the same as GLASS-JWST with the exception of the F090W filter, and overall shorter exposure times. One of the two NIRCam modules overlaps with UNCOVER. 

In Table~\ref{tab:exptime}, we list the exposure times adopted in the various filters for each of the aforementioned programs, while the footprints of the fields are illustrated in Figure~\ref{fig:mosaic}.

As a result of the overlap between programs and of their different observation strategies, such as the different PAs adopted that created multiple star diffraction spikes in the overlapping regions, the resulting exposure map is complex and inhomogenous across bands and area.
An analysis of the depth resulting from this exposure map is reported below.

\begin{deluxetable}{ccccc}\label{tab:exptime}
\tablecaption{NIRCam Exposure time}
\tablewidth{10pt}
\tablehead{\colhead{Filter} & \colhead{GLASS1} & \colhead{GLASS2} & \colhead{DDT1/2} & \colhead{UNCOVER1/2/3/4}}\startdata
F090W & 11520 & 16492 & - & - \\
F115W & 11520 & 16492 & 2104 & 10823\\
F150W & 6120 & 8246 & 2104 & 10823 \\
F200W & 5400 & 8246 & 2104 & 6700 \\
F277W & 5400 & 8246 & 2104 & 6700 \\
F356W & 6120 & 8246 & 2104 & 6700 \\
F410M & -    & -   & -  & 6700 \\
F444W & 23400 & 32983& 2104 & 8246 \\
\enddata
\tablecomments{Exposure time (in seconds) for each pointing of the three programs considered here.}
\end{deluxetable}

\subsection{Data reduction}
\subsubsection{Pre-reduction steps}

Image pre-reduction was executed using the official $JWST$ calibration pipeline, provided by the Space Telescope Science Institute (STScI) as a Python software suite\footnote{\url{https://jwst-pipeline.readthedocs.io/en/latest/}}. We adopted Version 1.8.2 \citep{Bushouse_JWST_Calibration_Pipeline_V182} of the pipeline and  Versions between \texttt{jwst\_1014.pmap} and \texttt{jwst\_1019.pmap} of the CRDS files (the only change between these versions is the astrometric calibration, that is dealt with as described below). We executed the first two stages of the pipeline (i.e. \texttt{calwebb\_detector1} and \texttt{calwebb\_image2}), adopting the optimized parameters for the NIRCam imaging mode,  which convert single detector raw images into photometric calibrated images. 

Using the first pipeline stage \texttt{calwebb\_detector1} we processed the raw uncalibrated data (\texttt{uncal.fits}) in order to apply detector-level corrections performed on a group-by-group basis. These include dark subtractions, reference pixels corrections, non-linearity corrections and jump detection that allows to identify cosmic rays (CR) events on the single groups. The last step of this pipeline stage allows us to derive the mean count rate, in units of counts per second, for each pixel by performing a linear fit to the data in the input image (the so-called ramp-fitting) excluding the group masked due to the identification of a cosmic ray jump. 

The output files of the previous steps (\texttt{rate.fits}) are processed through the second pipeline stage \texttt{calwebb\_image2}, which consists of additional instrument-level and observing-mode corrections and calibrations, as the geometric-distortion correction, the flat-fielding, and the photometric calibrations that convert the data from units of countrate to surface brightness (i.e., MJy per steradian) generate a fully calibrated exposure (\texttt{cal.fits}). 

The \texttt{cal.fits}  file also contains an \texttt{RMS} extension, which combines the contribution of all pixel noise sources, and a \texttt {DQ} mask where the first bit (\texttt{DO\_NOT\_USE}) identifies pixels that should not be used during the resampling phase.

We then applied a number of custom procedures to remove instrumental defects that are not dealt with the STScI pipeline. Some of them have already been adopted in \citep[][hereafter M22]{Merlin2022} and described there: we illustrate below only the major changes to the STScI pipeline in the default configuration and/or to the procedure adopted in M22.

\begin{figure}
\center
\includegraphics[width=0.49\textwidth]{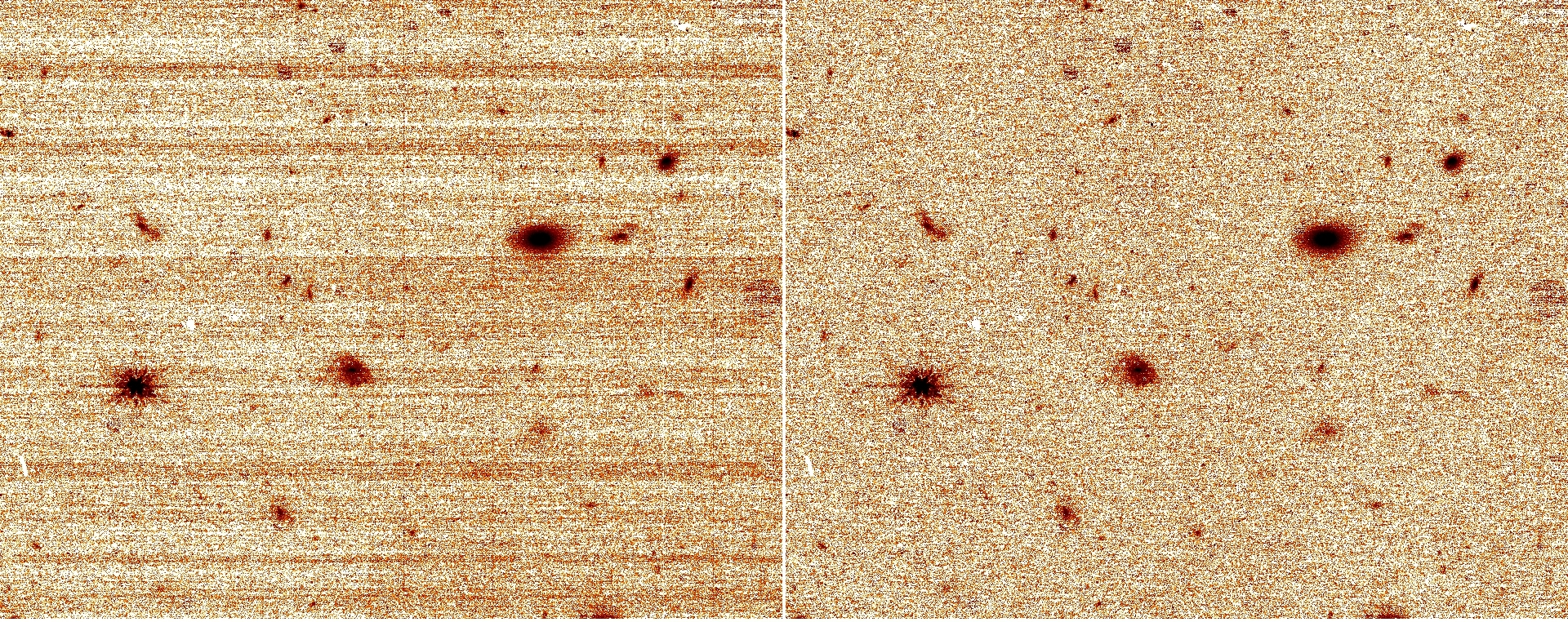}
\includegraphics[width=0.49\textwidth]{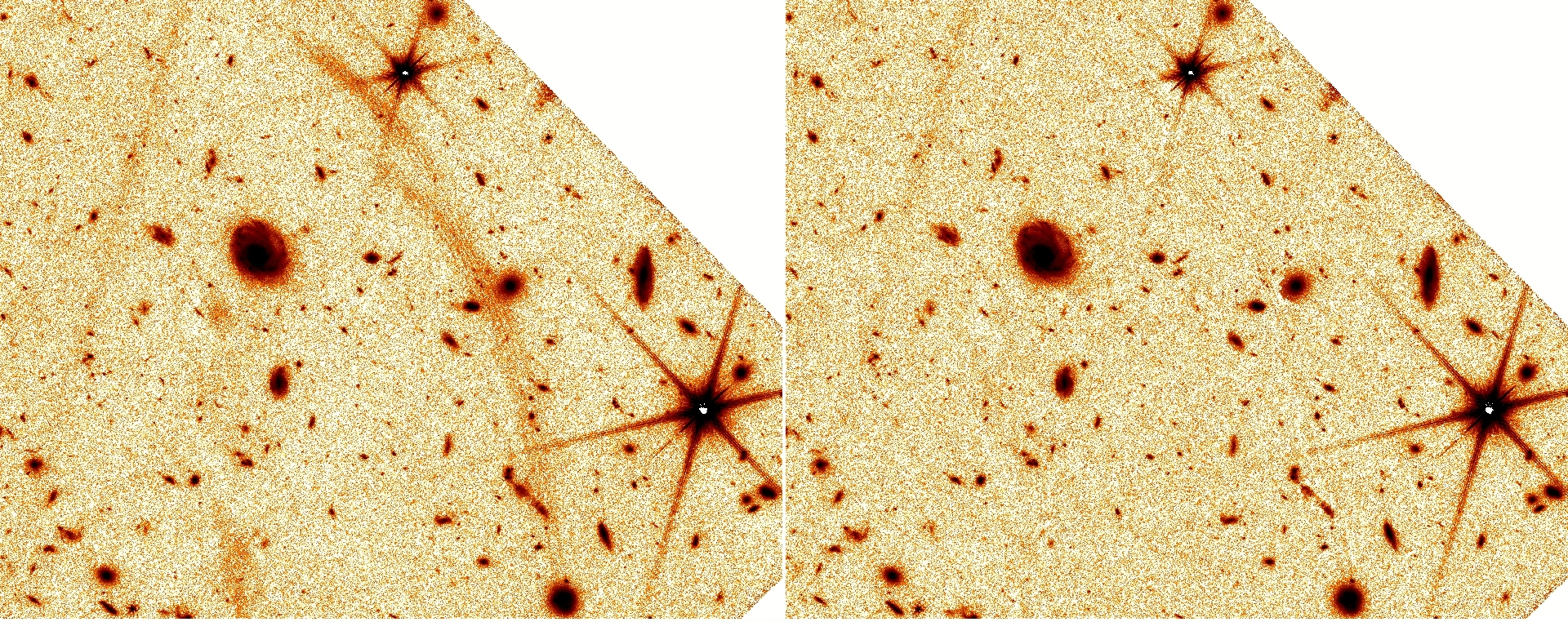}
\caption{Examples of custom procedures to remove residual instrumental defects not dealt with the current STScI pipeline. \textit{Top}: 1/f stripes removal on a GLASS F200W single exposure. \textit{Bottom}: A portion of the GLASS F150W mosaic before and after the claws treatment. 
}
 \label{fig:claws}
\end{figure}

\begin{itemize}

\item \textit{``Snowballs''}, i.e. circular artifacts observed in the in-flight data caused by a large cosmic ray impacts. Those hits leave a bright ring-shaped defect in the image since the affected pixels are just partially identified and masked. In M22, we developed a technique to fully mask out these features, which was not necessary here. Indeed, version 1.8.1 of the $JWST$ pipeline introduced the option to identify snowball events, expanding the typical  masking area to include all the pixels affected. This new implementation provides the opportunity to correct these artifacts directly at the ramp fitting stage, at the cost of a larger noise on the corresponding pixels. We enabled this non-default option,  and fine tuned the corresponding parameters to completely mask all the observed snowballs and, at the same time, minimize the size of high noise areas. 

\item \textit{``NL Mask''}: we find groups of deviant bright pixels on the \texttt{cal} images taken with the NIRCam Module B LW detector, more evident on deeper pointings. They result as not well corrected during the pre-reduction stage and are identified as ``WELL\_NOT\_DEFINED'' pixels\footnote{\url{https://www.stsci.edu/files/live/sites/www/files/home/jwst/documentation/technical-documents/_documents/JWST-STScI-004714.pdf}} in the Non Linearity Calibration file \footnote{\url{https://jwst-crds.stsci.edu/browse/jwst\_nircam\_linearity\_0011.rmap}}. We recognize them by their flag in the DQ and  mark them as \texttt{DO\_NOT\_USE} for the coaddition.

\item \textit{1/f noise}, which introduces random vertical and horizontal stripes into the images \citep[see][]{Schlawin2020}. We remove this by subtracting the median value from each line/column. To remove the flux from objects as accurately as possible, we  mask out all objects and the bad pixels flagged in the data quality. The masks were obtained from the segmentation maps obtained with \textsc{SExtractor} (Version 2.25.0) \citep{Bertin1996} and then they were further dilated in order to exclude the contamination from the faint outskirt of the objects, which escape detection below the SExtractor  threshold. We have applied a differential procedure to dilate objects depending on their ISOAREA: the segmentation of objects with ISOAREA$<$5000 pixels was dilated using a $3\times 3$ convolution kernel and a dilation of 15 pixels, while for the segmentation of objects with ISOAREA$\geqslant$5000 pixels a $9\times 9$ convolution kernel and a dilation of $4\times15$ pixels was used. The procedure was executed separately for each amplifier in the SW detectors (i.e. 4 times for each individual image) with the exception of the denser areas corresponding to the centers of the clusters and the brightest field star, where objects are significantly larger than the amplifier width (500 pixels, corresponding to about 30'') and could not be masked efficiently. In this case we removed the $1/f$ noise over the entire row. Our procedure, which was already adopted on the first release of the GLASS data (M22), is conceptually similar to the one adopted for the CEERS data \citep{Bagley2022}.  We publicly release the code adopted for this step.

\item \textit{Scattered light}: we identify additive features in the F115W, F150W and F200W images. These low-surface brightness features have already been revealed by commissioning data \citep[see][]{Rigby2022} and are due to scattered light entering the optical path. These anomalies have been dubbed \textit{wisps} or \textit{claws}, depending on their origin and morphology. \textit{Wisps} have a nearly constant shape and can thus be subtracted from the images with simple templates. We removed these features by extracting their 2D profile from the available template (we do not use the entire template image to avoid subtracting its empty but noisy regions) and then normalizing the residual template to match the feature intensity in each image. \textit{Claws} have been first identified and singled out in images. Their shape on each image has been reconstructed by interpolating a 2D mesh with box size 32 pixels and then eventually subtracted from the same image. These procedures efficiently remove most of these features, as shown in Figure~\ref{fig:claws}. 

We found other defects in the F090W image, and to a lesser extent in the F115W, on images taken on June 2022. These defects consist of additional scattered light in the images, resulting in artificial sources along the FoV and they are due to a so-called ``wing-tilt event'', i.e., a small shift of one of the wings of the primary mirror. We adopted the procedure described in M22 to identify and mask these artifacts from images. We emphasize that they do not affect images taken on November 2022.

\end{itemize}

We then re-scaled the single exposures to units of $\mu$Jy/pixel, using the conversion factors output by the pipeline. 

We also note that our procedure to remove the $1/f$ noise and the background effectively removes  the intra-cluster light (ICL) from the images. We caution the user to avoid using these images to study in the detail the ICL.

\subsubsection{Astrometry}
\begin{figure*}
\center
\includegraphics[width=0.195\textwidth,height=0.3\textwidth]{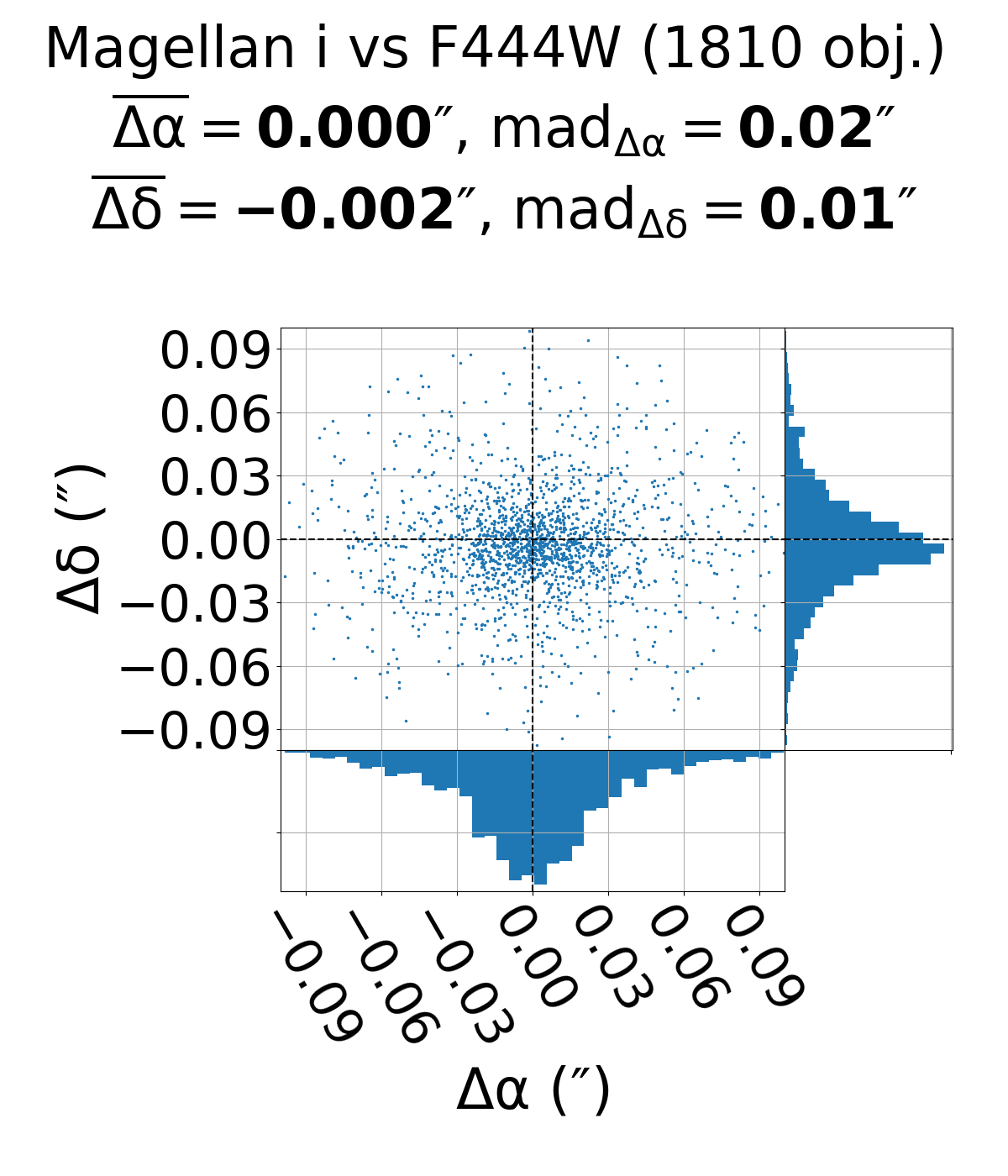}
\includegraphics[width=0.195\textwidth,height=0.3\textwidth]{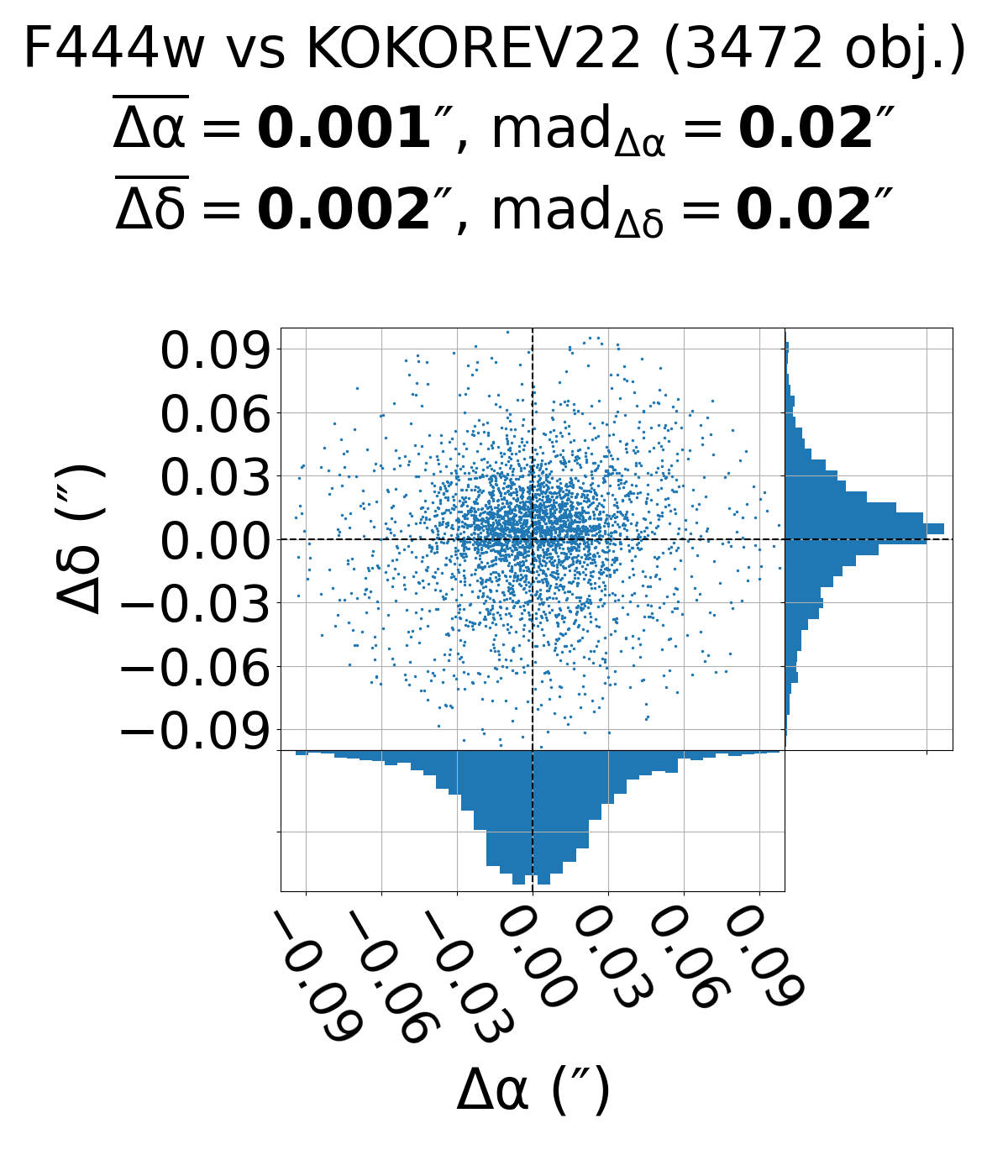}
\includegraphics[width=0.195\textwidth,height=0.3\textwidth]{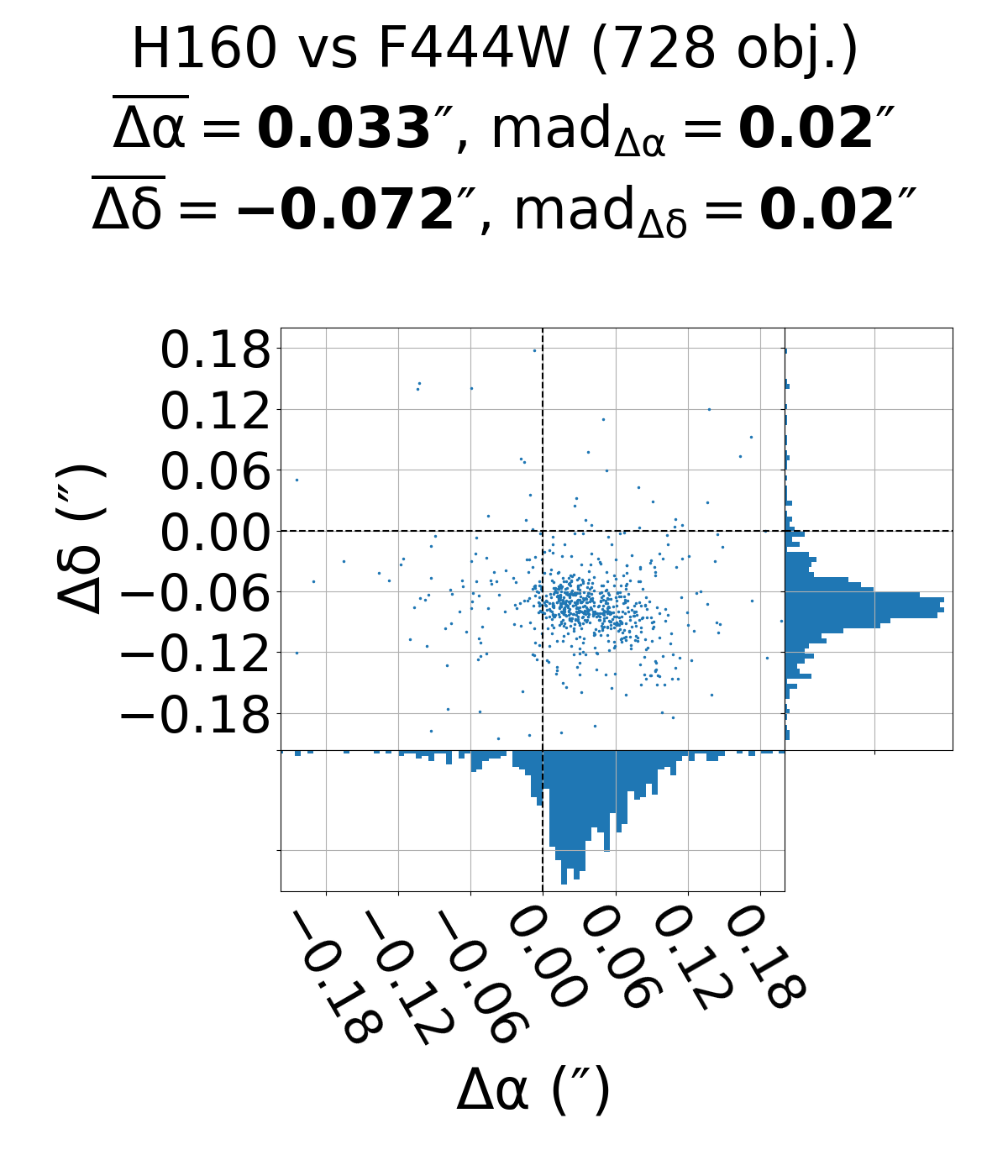}
\includegraphics[width=0.195\textwidth,height=0.3\textwidth]{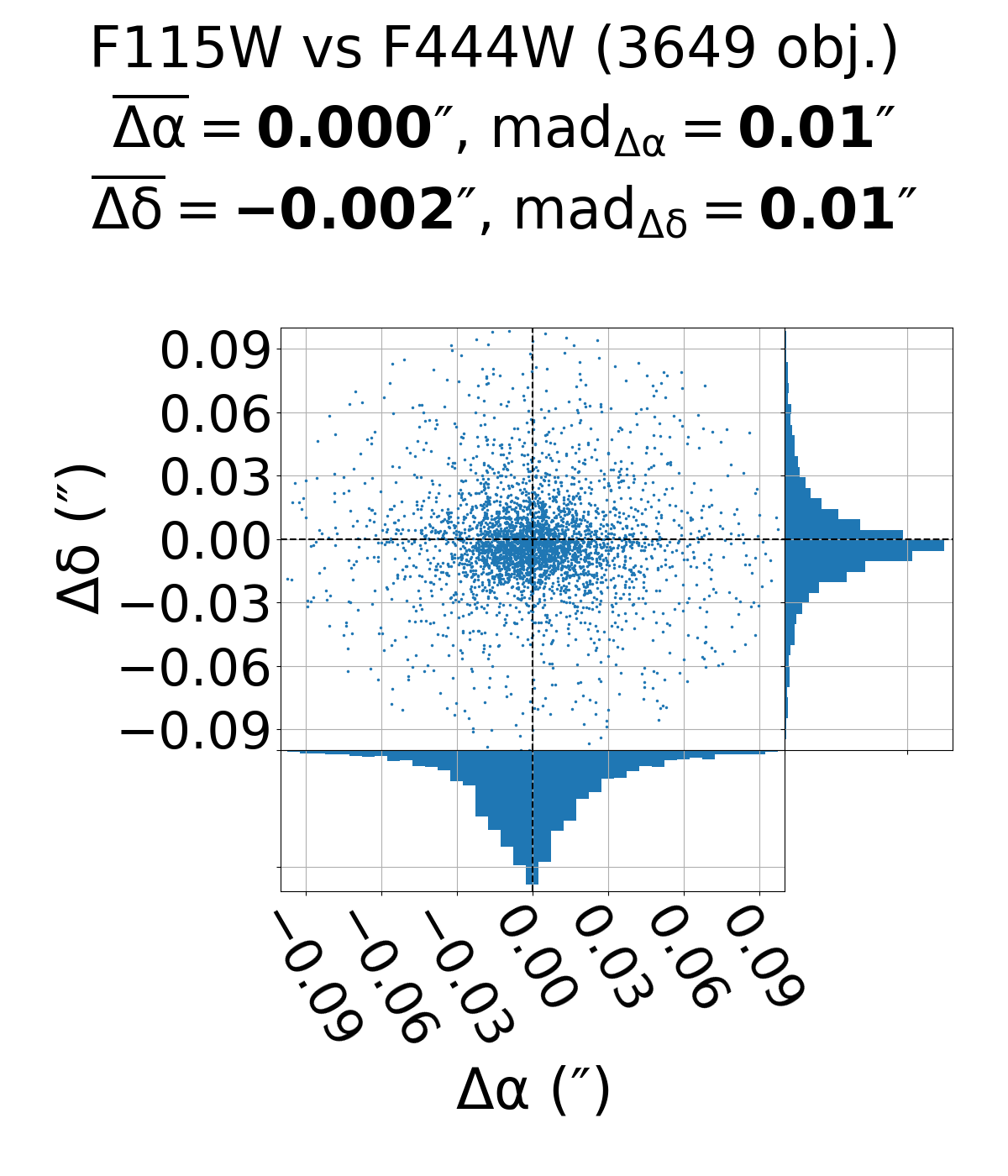}
\includegraphics[width=0.195\textwidth,height=0.3\textwidth]{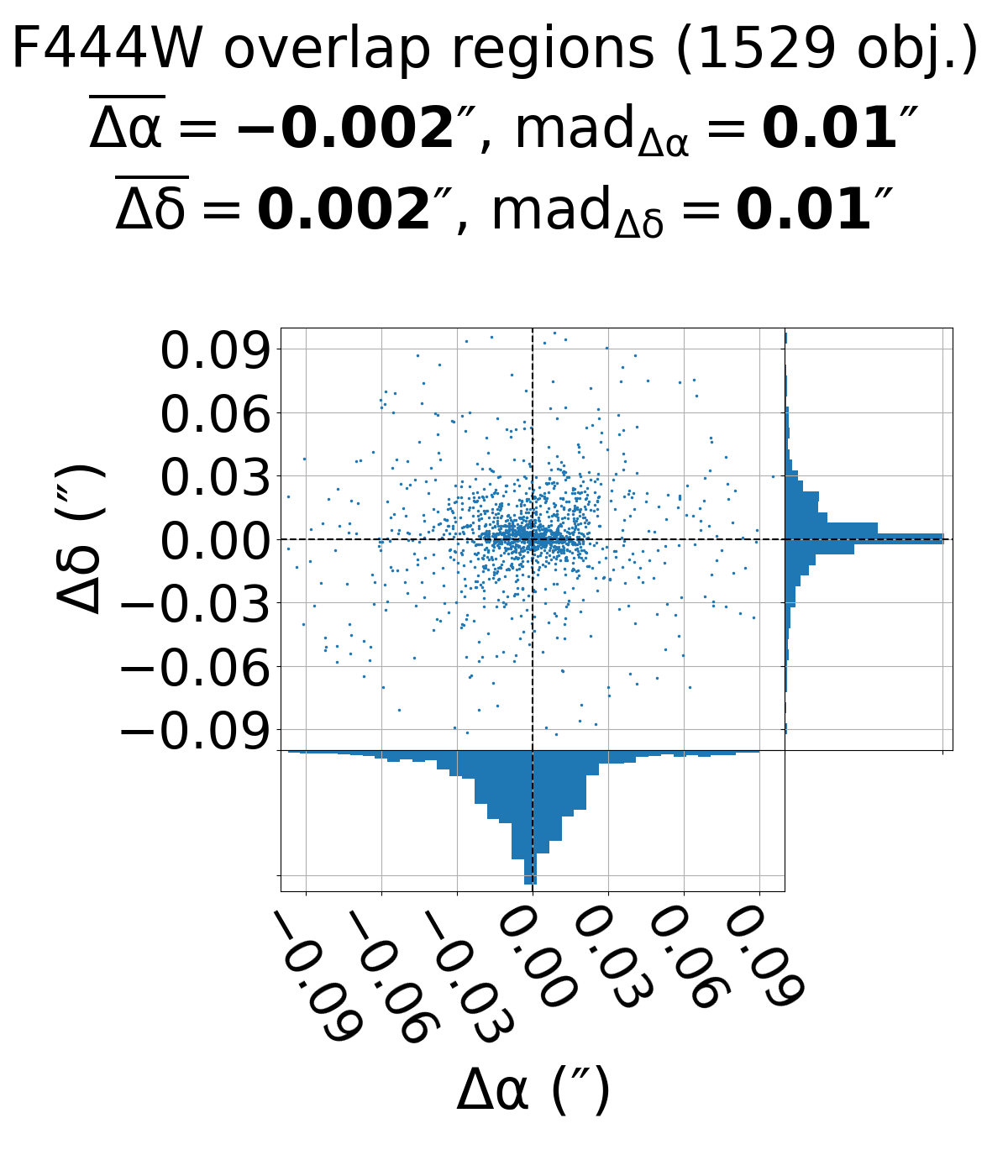}
\caption{Validation tests on the astrometric registration.
The scatter diagrams show the displacement $\delta RA$ and $\delta DEC$  of sources between the sources detected in this catalog and those in several reference catalogs, namely:
\textit{Left}: the Magellan $i$--band catalog registered to Gaia DR3 used as global reference for calibration. \textit{Middle left}: the catalog released by \citep{Kokorev22}. \textit{Middle}: the AstroDeep catalog on the central region of the A2744 cluster \citep{Merlin2016b}. \textit{Middle right}: The catalog of the F115W-detected sources in this images. The \textit{Right} plot shows the objects detected on UNCOVER--only images and those in the GLASS and DDT samples on two overlapping regions. 
In all diagrams the average value $\Delta \alpha$ and $\Delta \delta$ and the median average deviation mad$_{\Delta \alpha}$ and mad$_{\Delta \delta}$ are reported.
}
 \label{fig:astrom}
\end{figure*}

The astrometric calibration was performed using \textsc{SCAMP} \citep[][]{Bertin2006}, with third-order distortion corrections. Compared to the procedure we adopted in M22, we started from the distortion coefficient computed by the STScI pipeline, stored in the \texttt{cal} images. We refined the astrometric solution by running \textsc{scamp} in \texttt{cal}  mode, which optimizes the solution with limited variations from the starting solution. We have found this procedure both accurate and reliable, as described below. We first obtained a global astrometric solution for the F444W image, which is usually the deepest. As there not enough GAIA-DR3 stars usable for every NIRCam detector, we have aligned the images to a ground-based catalog obtained in the $i$-band with the Magellan telescope in good seeing condition (see T22 for details) of the same region, which had been previously aligned to GAIA-DR3 stars \citep[][2022 in prep.]{Prusti2016}. We then took the resulting high-resolution catalog in F444W as reference for the other $JWST$ bands, using compact, isolated sources detected at high signal-to-noise at all wavelengths. Each NIRCam detector has been analysed independently, in order to simplify the treatment of distortions and minimise the offsets of the sources in different exposures. Finally, we used \textsc{SWarp} \citep[][]{Bertin2002} to combine the single exposures into mosaics projected onto a common aligned grid of pixels, and \textsc{SExtractor} to further clean the images by subtracting the residual sky background. The pixel scale of all the images was set to 0.031$\arcsec$ (the approximate native value of the short wavelength bands), to allow for simple processing with photometric algorithms. 

The final image, computed as a weighted stack of all the images from the three programs, has a size of $24397\times21040$ pixels, corresponding to $12.6\times10.87$~arcmin$^2$. In this frame, the area covered by the wide-band NIRCam images (F115W, F150W, F200W, F277W, F356W and F444W) is of exactly 46.5 arcmin$^2$. 

Given the especially deep and sharp nature of the $JWST$ images, where most of the faint objects have sizes below 0.5$\arcsec$, the requirements on the final astrometric accuracy are extremely tight, to avoid errors in the multi-band photometry (where a displacement of as little as 0.1$\arcsec$ can bias color estimates). These requirements must be met also in the overlapping regions of the various surveys, which have often been observed with different detectors. 

To verify the final astrometric solution we conducted a number of validation tests, where we compare the positions of cross-matched objects in catalogs extracted from different images. For each of these catalogs we used \textsc{SExtractor} in single image mode and adopted the \textsc{XWIN} and \textsc{YWIN} estimators of the object center, which are more accurate than other choices. Given the unprecedented image quality of NIRCam, the center of extra--galactic objects with complex morphology may be difficult to estimate with high accuracy, particularly when observed across a large wavelength interval. Therefore we  only compared objects with well--defined positions to minimize errors, using the $\Delta X, \Delta Y =$\textsc{ERRAWIN\_WORLD, ERRBWIN\_WORLD} estimators of the error and limited the analysis to objects with $(\Delta X^2+\Delta Y^2)^{1/2} \leq 0.018\arcsec$. 
From these catalogs we estimated both the average offset of the object centers  $\overline{\Delta \alpha}$ and $\overline{\Delta \delta}$, and the median average deviation mad$_{\alpha}$ and mad$_{\delta}$, which measure the intrinsic scatter in the alignment. In Figure~\ref{fig:astrom} we report the main outcome of these tests:
\begin{itemize}
    \item (\textit{Left}) We first compared the positions of objects in the original Magellan $i$-band and the resulting F444W of the entire mosaic. We find an essentially zero offset and mad$_{\alpha} \simeq$~mad$_{\delta}\simeq 0.02\arcsec$, which is $2/3$ of a pixel. 
    \item (\textit{Middle left}) We compared the F444W catalog with the catalog released by \citep{Kokorev22} in the context of the ALMA lensing cluster survey (ALCS), containing HST and IRAC sources in the A2744 region. Overall, the comparison results in a good alignment with a very small offset ($\overline{\Delta \alpha} \simeq 1 mas$ and $\overline{\Delta \delta} \simeq 2 mas$) and mad$_{\alpha}$ $\simeq$ mad$_{\delta}\simeq 0.02\arcsec$    
    \item (\textit{Middle}) We compared the F444W catalog with the 
    AstroDeep $H_{160}$ catalog obtained on the central region of the A2744 cluster, as obtained in the context of the Frontier Fields initiative \citep{Merlin2016b}. 
    While the intrinsec scatter is still good (mad$_{\alpha}$ $\simeq$ mad$_{\delta}\simeq 0.02\arcsec$), we find a systematic offset by about 1 pixel in RA and 2.5 pixels in DEC, which is most likely due to different choices in the absolute calibration of the ACS/WFC3 data released within the Frontier Fields.
    \item (\textit{Middle right}) We compare here the relative calibration of filters at the two extremes of the spectral range, F444W and F115W, where morphological variations and color terms may change the center position and affect the astrometric procedure. We find again very good alignment with negligible offset and small mad$_{\alpha}$ $\simeq$ mad$_{\delta}\simeq 0.01\arcsec$.
    \item (\textit{Right}) Finally, we compare the astrometric solutions on the overlapping areas by summing independently the data of the three different programs and checking the accuracy in the overlapping area. Again we find very good alignment with negligible offset and small mad$_{\alpha}$ $\simeq$ mad$_{\delta}\simeq 0.01\arcsec$.

\end{itemize}

We adopted a cross matching radius $r=0.1\arcsec$ in most of the validation tests, except in comparison with the AstroDeep catalog, for which we adopted a much larger radius  ($r=0.4\arcsec$), given the larger offset at level of $0.075\arcsec$.  We also note that the scatter in $\delta$ seems consistently lower than that in $\alpha$, but we failed to identify a clear origin for this effect, that does not impact the global accuracy.
We therefore conclude that the astrometric procedure is accurate and adequate for the goals of this Stage I release. In the future we plan to explore further and validate other options for astrometric registration and also release images with a smaller pixel scale, to better exploit the unprecedented image quality of the $JWST$ data. However, we note that the GLASS-$JWST$ data have a very limited dithering pattern (which was driven by spectroscopic requirements) and so may benefit only marginally from moving to smaller pixels.

\subsection{Estimating the Final Depth}
\begin{figure}
\center
\includegraphics[width=0.49\textwidth]{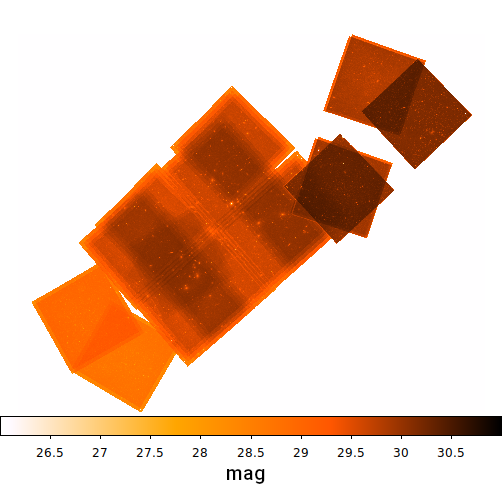}
\caption{Depth of the full mosaic F444W image, as produced by our pipeline based on the variance image of each exposure and with the re-normalization described in the text. Each pixel has been converted into $5\sigma$ limiting flux computed on a circular aperture of 0.2''.
}
 \label{fig:depth}
\end{figure}
The final coaddition of the different images is weighted according to their depth, as estimated by the RMS image produced by the pipeline. We therefore obtain an optimally averaged image with the resulting RMS image. We \textit{a posteriori} verified whether the noise estimate encoded in the 
RMS effectively reproduced the photometric noise.
To do this, we injected artificial point sources of known magnitude in empty regions of the image and measured their fluxes and uncertainties with \textsc{a-phot} \citep{Merlin2019}, using apertures of radius 0.1\arcsec. To consider that the mosaics have varying depths resulting from a complex pattern of different exposures, we perfomed this analysis separately over four different image regions, chosen to have approximately constant exposure time.

In general, we find that the RMS of the resulting flux distribution is 1.1$\times$ larger than the value we would expect from the \textsc{SExtractor} errors, computed from the RMS image. Furthermore, a larger difference (1.4$\times$) is found for the F444W GLASS image, which is affected by a residual pattern due to poor flat--fielding with the current calibration data. We therefore re-scaled the RMS maps produced by the pipeline according to these factors.

The resulting depth of this procedure is shown in Figure~\ref{fig:depth}. The RMS image is converted into a $5\sigma$ limiting flux computed on a circular aperture with a diameter of 0.2$\arcsec$, which is the size adopted to estimate colors of faint sources. The depth ranges from $\simeq 28.6$ AB on the DDT2 footprint (in particular the area not overlapping with DDT1) to $\simeq 30.2$ AB in the  area where GLASS1 and GLASS2 overlap, arguably one of the deepest images obtained so far by $JWST$.

\begin{figure}
\center
\includegraphics[width=0.45\textwidth]{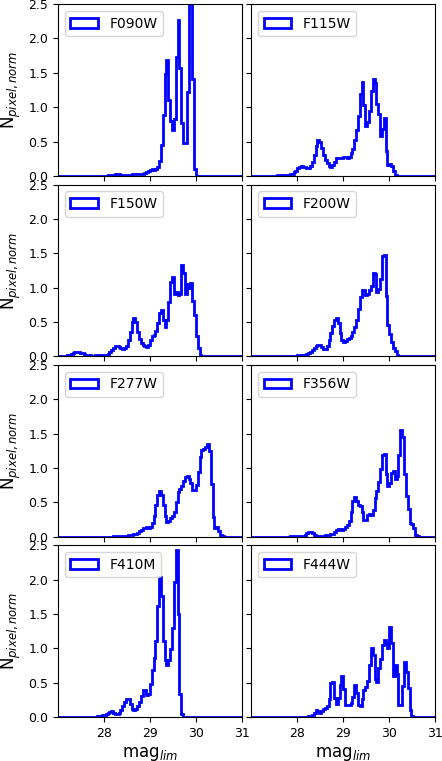}
\caption{Distribution of the limiting magnitude for each band, as shown in the legend. Limiting magnitudes per pixel are computed as in Figure~\ref{fig:depth}, converting each rms pixel into the corresponding $5\sigma$ limiting flux computed on a circular aperture of $0.2\arcsec$}. 
 \label{fig:depthhist}
\end{figure}

A more quantitative assessment of the depth in the various filters is reported in Figure~\ref{fig:depthhist}, where we show the distribution of the limiting magnitudes in each image resulting from the different strategies adopted by the surveys,computed as described above. A clear pattern is seen, illustrating the large, mid--depth area obtained by UNCOVER and the shallower and deeper parts obtained by DDT and GLASS respectively.

\subsection{$HST$ Imaging}
We have also used the existing images obtained with $HST$ in previous programs, namely with the  F435W, F606W, F775W and F814W  bands with ACS and the F105W, F125W, F140W and F160W bands with WFC3 - other $HST$ data are available from MAST but are either too shallow and/or limited in area and are not used in our work. Among these data are included also the images that we obtained with DDT Program HST-GO-17231 (PI: Treu), which was specifically aimed at obtaining ACS coverage for the majority of the GLASS1 and GLASS 2 fields. 
We have used calibrated stacked image and weights (G. Brammer, private communication) 
that we  have realigned (after checking that the astrometric solution is consistent) onto our reference grid to allow a straightforward computation of colors.

\section{Photometric Catalog}
\label{sec:catalog}
\subsection{Detection}
\label{sec:detection}

We follow here the same prescriptions adopted by M22 and \citet{Castellano2022b,Castellano2023}. We performed source detections on the F444W band, since it is generally the deepest or among the deepest image for each data set, and because
high-redshift sources (which are the main focus of these observations) are typically brighter at longer wavelengths, where they are observed beyond the Balmer break and in a region dominated by emission lines. This approach has the advantage of delivering a clear-cut criterion for the object detections, that can easily be translated into a cut of rest-frame properties for high redshift sources.

We used \textsc{SExtractor}, adopting a double--pass object detection as applied  for the HST-CANDELS campaign \citep[see][]{Galametz2013}, to detect the objects, following the recipes and parameters described in M22. We note, in particular, that we adopt a detection threshold corresponding to a signal-to-noise ratio (SNR) of 2. This is based on simulations and on the visual inspection of the images at various wavelengths, and has been chosen  to maximise the number of detected sources while maintaining the number of spurious ones still limited, as discussed in M22.  As in M22, we adopted the following \textsc{SExtractor} parameters: \texttt{DETECT\_MINAREA}$=8$,   \texttt{DETECT\_THRESH}$=$\texttt{ANALYSIS\_THRESH}$=0.7071$, \texttt{DEBLEND\_NTHRESH}$=32$, \texttt{DEBLEND\_MINCOUNT}$=0.0003$, \texttt{BACK\_SIZE}$=64$, \texttt{BACK\_FILTERSIZE}$=3$, \texttt{CLEAN\_PARAM}$=1$ and detection has been performed adopting a gaussian filter with FWHM$=0.14\arcsec$.

The final \textsc{SExtractor} catalog on the entire A2744 area contains 24389 objects.

Estimating the completeness and purity in a non-contiguous (in terms of area and exposure) mosaic derived from the large number of observations adopted here, is intrinsically ambiguous. As shown in Figure~\ref{fig:depthhist} the depth of these images spans approximately 2 magnitudes, and the completeness is therefore inhomogenoues - not to mention the existence of the cluster that complicates both the detection and the estimate of the foreground volume (C22b). For these reasons, we do not attempt the traditional estimate of the completeness and refer to Figure~\ref{fig:depth} and to Figure~\ref{fig:depthhist} for an evaluation of the depth. For a proper analysis of the completeness we refer the reader to the methodology adopted by C22b where we estimate the completeness separately on the individual mosaics of the three data sets, which were processed independently. We make the three mosaics available upon request for this purpose.

\subsection{Photometry}
\label{sec:photometry}
We have compiled a multi-wavelength photometric catalog following again the prescriptions of M22, which in turn is based on previous experience with Hubble Space Telescope ($HST$) images in CANDELS \citep[see e.g.][]{Galametz2013} and in AstroDeep \citep{Merlin2016b, Merlin2021}. The catalog is based on a detection performed on the F444W image described above, and PSF--matched aperture photometry of all the sources. We include all the NIRCam images presented here and existing images obtained with $HST$ in previous programs, namely with the  F435W, F606W, F775W and F814W  bands with ACS and the F105W, F125W, F140W and F160W bands with WFC3.

The images considered here have PSFs that range from 0.035'' to 0.2''. Considering that most of the objects have small sizes, with half--light--radii less than 0.2'', it is necessary to apply a PSF homogenization to avoid bias in the derivation of color across the spectral range.

\subsubsection{PSF matching}
 Since the detection band has the coarsest resolution, we PSF-matched all the other NIRCam images to it for color fidelity. We created convolution kernels using 
the \textsc{WebbPSF} models publicly provided by STScI\footnote{\url{https://jwst-docs.stsci.edu/jwst-near-infrared-camera/nircam-predicted-performance/nircam-point-spread-functions}}, combining them with a Wiener filtering algorithm based on the one described in \citet{Boucaud2016}. To smooth the images, we used a customised version of the convolution module in \textsc{t-phot} \citep{Merlin2015,Merlin2016a}, which uses \texttt{FFTW3} libraries. However, we note that this approach cannot fully correct for the inhomogeneities of the PSF:  the calibration upon which \textsc{WebbPSF} is calibrated is inevitably initial, and the $JWST$ PSF is  time-- and position--dependent \citep{Nardiello2022}, and our dataset is the inhomogeneous combination of data obtained at different times and with different PA, so that the PSF definitely changes over the field. For this version of the catalog we used the UNCOVER PSF models (epoch: 2022/11/07, PA: 41.2 deg) as average PSFs, and we plan to improve our PSF estimation in the future versions of the catalog that will be released in Stage II.

Similarly, concerning the $HST$ images, we note that all of them have too few stars to obtain a robust estimate of the PSF directly from the images, so we adopt, in all cases, existing $HST$ PSFs, taken from CANDELS. This approximation may introduce small biases in the final catalog.  ACS images have  been PSF-matched to F444W, while for  the WFC3 F105W, F125W, F140W and F160W images, which have a PSF larger than the F444W one, we have done the inverse - smoothed  the F444W image and the WFC3 F105W, F125W, F140W to the F160W and followed a slightly different procedure that we describe below. 

\subsubsection{Flux estimate}
The total flux is measured with \textsc{a-phot} on the detection image F444W  by means of a Kron elliptical aperture \citep{Kron1980}. As we have shown in M22, simulations suggest that Kron fluxes measured with \textsc{a-phot} are are somewhat less affected by systematic errors, while being slightly more noisy.

Then, we used \textsc{a-phot} to measure the fluxes at the positions of the detected sources on the PSF-matched images, masking neighboring objects using the \textsc{SExtractor} segmentation map. Given the wide range of magnitudes and sizes of the target galaxies we have measured the flux in a range of apertures. The segmentation area (the images being on the same grid and PSF-matched) and five circular apertures with diameters that are integer multiples ($2\times, 3\times, 8\times, 16\times$) of the  FWHM in the F444W band, that correspond to  0.28$\arcsec$, 0.42$\arcsec$, 1.12$\arcsec$ and 2.24$\arcsec$ diameters. 
For the four WFC3 images (which have a PSF larger than F444W) we first filtered the F444W to their FWHM and then measured colors between the filtered F444W and the WFC3 images.  To minimize biases when these colors are combined with those of the other bands, we use in this case apertures the same multiples of the WFC3 PSF adopted for the other bands. We remark that this procedure is only approximate, and delivers a first order correction of the systematic effects due to different PSFs. 
In a future release we plan to adopt more sophisticated approaches to optimize photometry, including but not limited to the improvement of the PSF estimate and applying \textsc{t-phot} on WFC3 images that have a larger PSF.

Total fluxes are obtained in the other bands by normalizing the colors in a given aperture to the F444W total flux, i.e. by computing $f_{m,total} = f_{m,aper}/f_{F444W,aper} \times f_{F444W,total}$, as described in M22.

We release the five catalogs described above (one computed on segmentation and four on the different apertures) and we leave the user to choose which is the most suitable for a given science application. In general small-aperture catalogs are more appropriate for faint sources as they match their small sizes and minimize the effect of contamination from nearby sources, that can be important for blended objects. Larger apertures may be more appropriate for brighter sources and especially cluster members.

\subsubsection{Validation tests}

\begin{figure}
\center
\includegraphics[width=0.49\textwidth]{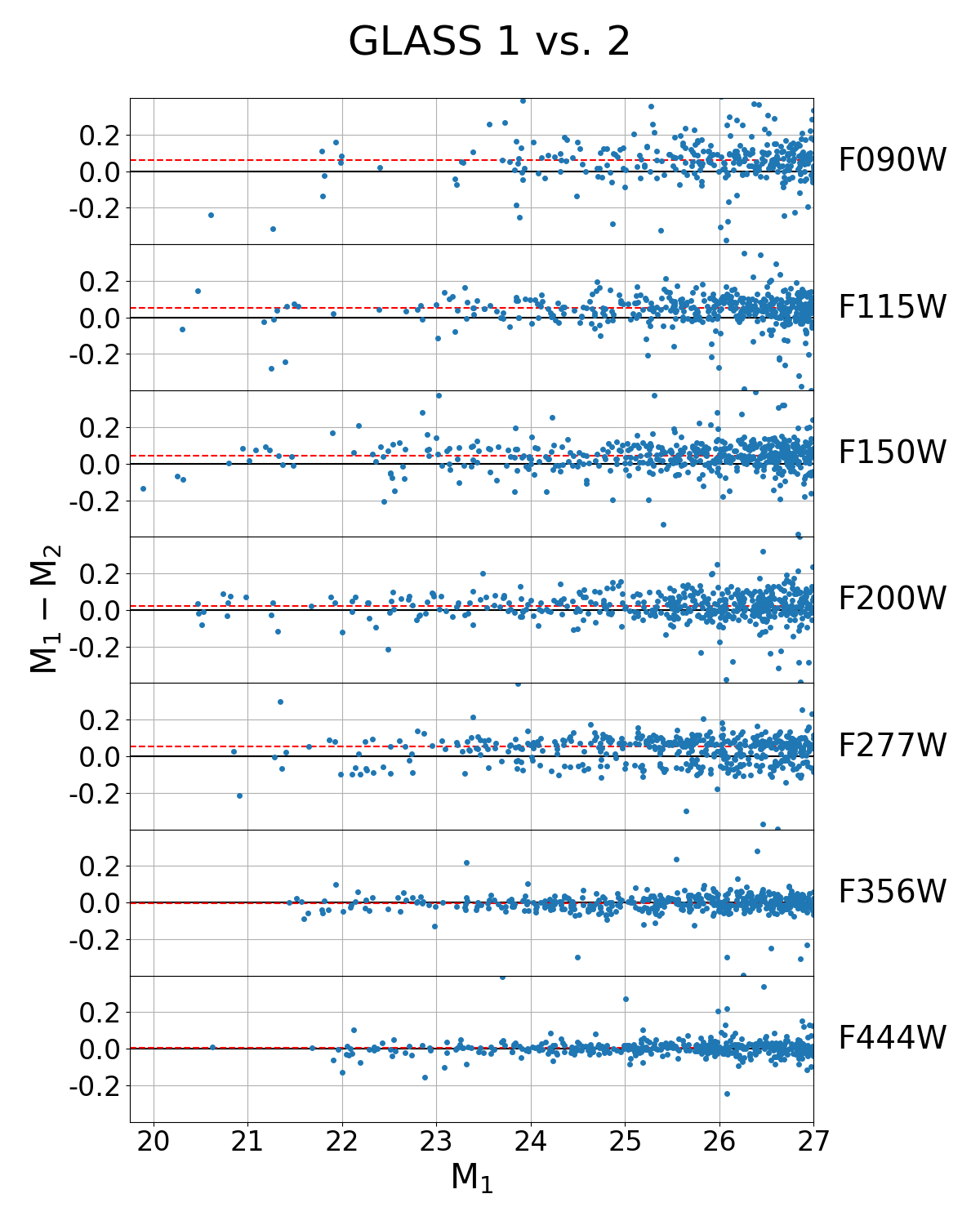}
\caption{Stability of the photometric calibration between different detectors, as measured by comparing the photometry of high S/N objects ($S/N>25$) detected in the two epochs of observations in the SE quadrant of GLASS (lower leftmost green square in Figure~\ref{fig:mosaic}). Objects in this area have been observed in two epochs (June and November 2022) and with modules B and A, respectively. For each filter difference in magnitude $\Delta M = M_{1} - M_{2}$ for objects between epoch1 and epoch2  as a function of $M_{1}$ is reported. 
Red dashed lines represent the median offsets, namely we found: $\overline{\Delta M} \approx 0.06$ with $mad \approx 0.05$ for F090W,  $\overline{\Delta M} \approx 0.05$ with  $mad \approx 0.04$ for F115W,  $\overline{\Delta M} \approx 0.04$ with  $mad \approx 0.04$ for F150W,  $\overline{\Delta M} \approx 0.02$ with  $mad \approx 0.04$ for F200W,  $\overline{\Delta M} \approx 0.05$ with  $mad \approx 0.04$ for F277W, 
 and negligible  in F356W and F444W with $mad \approx 0.03$ and  $mad \approx 0.02$ respectively. We have visually inspected the bright objects with $|\Delta M| > 0.05$ and verified that they mostly originate from saturated stars or objects with incomplete coverage. 
}
\label{fig:DM}
\end{figure}
We have performed a few validation tests to verify primarily the flux calibration, which has been the subject of many revisions in these first months, and to a lesser extent, of the procedure adopted to derive the photometric catalog.

The overlap between GLASS1 and GLASS2 southern quadrants offers a nice opportunity to test the NIRCam flux calibration.
Indeed, the two GLASS observations have been observed in two epochs (June and November 2022) with a PA difference of nearly 150 degrees. As a result, the southern quadrant of GLASS1 and GLASS2 largely overlap but have been observed with modules B and A, respectively. We have therefore obtained stacked images of the two epochs separately, built a photometric catalog with the same recipes and checked the magnitude difference between objects observed with different detectors. The result of this exercise, done on all bands, is reported in Figure~\ref{fig:DM}. We note that in the short bands the two modules are made of 4 detectors, each with an independent calibration, which we plot all together in Figure~\ref{fig:DM}. The comparison, that is limited to objects observed with  high $S/N>25$, shows that the average magnitude difference between the two modules is in general quite small, in all cases below 0.05 \textit{mags} (see Figure~\ref{fig:DM} and its captions for details). This confirms that the flux calibration between the different modules is reasonably stable at this stage.

\begin{figure}
\center
\includegraphics[width=0.46\textwidth]{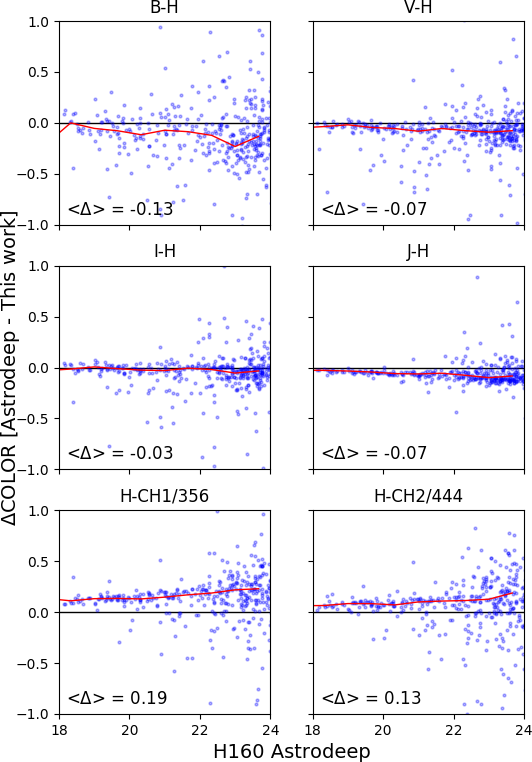}
\caption{Comparison between the colors measured in six bands from this catalog and from the Astrodeep \citep{Merlin2016a} catalog. Each panel show the difference in a color as a function of the H160 magnitude from the Astrodeep catalog. The red lines show the running medians and their weighted averages are also given.}
 \label{fig:astrodeep}
\end{figure}

As a further check to validate the photometric pipeline, we have compared the colors of the sources in the region of the A2744 cluster with those measured for the Frontier Fields survey \citep{Lotz2014} within the AstroDeep project \citep{Merlin2016b,Castellano2016}. We choose to compare colors of sources to avoid possible systematics deriving from the total fluxes estimates obtained on two different bands in the two catalogs; some residual offsets and trends can be due to the different segmented areas. 
The cross-match was performed selecting objects having $m_{H160}<24$, assuming a positional accuracy $\Delta r < 0.4\arcsec$ to account for the possible mismatch in absolute astrometry between the two. This comparison is shown in Figure~\ref{fig:astrodeep}, where we show isophotal colors computed on PSF-matched images as a function of the $H160$ magnitude of the Astrodeep catalog also providing running and global median offsets. 
We note that in \cite{Merlin2016b} we have explicitly modeled and subtracted the Intra Cluster Light (ICL) and the brightest cluster sources. This was especially needed to derive accurate colors on Spitzer images, that have a much poorer resolution. In this case, our procedure to remove the $1/f$ noise and the background effectively removes a significant fraction  of the ICL, to the point that objects falling in these are do not show any systematic shift of the objects in Figure~\ref{fig:astrodeep}. Therefore, the removal of the brightest cluster members is postponed to Stage II release.  We also note that, for the same reason, our images are not adequate to study the ICL emission. The comparison shows that  - when the same approach is used to estimate colors, i.e.  isophotal magnitudes are adopted - the agreement between the two catalogs is good (although the shallowness and low resolution of the IRAC bands makes the comparison less accurate).

From this comparison we conclude that - quite reassuringly - the overall photometric chain is consistent between the well established Frontier Fields data and these new data. At the same time, we remark that the optimal choice of the aperture depends on the size and kind of objects under study. Small apertures tend to have higher S/N and should be preferred for faint sources. For the brightest sources, larger apertures should be preferred. It is also possible to estimate rough color gradients by comparing the various apertures that we release. 
We also tested that applying the same technique without PSF matching introduces an offset of the order of $\sim$0.2 mags in the final colors, which would clearly affect the derived photometric redshifts and SED fitting results.

Finally, in an effort to cross-validate our results prior to release, in the lead up to this paper we compared our catalogs to those under development by the UNCOVER team (\cite{weaver2023uncover} in prep) based on the same raw datasets. The image processing and photometric procedures adopted by the two teams have significant differences. The main  are: i) image coaddition  (UNCOVER team adopts \textsc{Grizli} \citep{Brammer2019}, while we use a custom pipeline which uses \textsc{scamp} and \textsc{swarp}; ii) object detection (UNCOVER uses an optimally stacked F277W$+$F356W$+$F444W image after removing the intracluster light, while we use F444W); iii) techniques and tools for PSF matching and photometry.  For these reasons, we found some differences between the catalogs, especially for faint sources at the detection limit, as expected. However, our comparison of working versions of the catalogs produced by the two teams shows overall a good agreement in the colors and magnitudes of the vast majority of objects, 
as shown in Figure~\ref{fig:catglvsunc} in the Appendix.

\section{Summary}
\label{sec:conclusions}

We present the data obtained by three NIRCam programs on the A2744 cluster in this paper: the GLASS-JWST Early Release Science Program,  UNCOVER, and Directory Discretionary Time 2756. 
All the data, taken with eight different filters (F090W, F115W, F150W, F200W, F277W, F356W, F410M, F444W), have been reduced with an updated pipelines that builds upon the official STScI pipeline but includes a number of improvement to better remove some instrumental signature and streamline the process.

All frames have been aligned onto a common frame with a $0.031\arcsec$ pixel scale, approximately matching the native pixel scale of the short wavelength data.
The final images on the whole A2744 region cover an area of 46.5 arcmin$^2$ with PSF ranging from $0.035\arcsec$ (for the F090W image) to $0.14\arcsec$ (F444W), and reach astonishingly deep $5\sigma$ magnitude limits from 28.5 to 30.5, depending on location and filter. 

We exploit also other $HST$ publicly available programs which have targeted the area, including also the available $HST$ ACS and WFC3 data in the F435W, F606W, F775W and F814W (ACS)  and F105W, F125W, F140W and F160W (WFC3) bands, to expand the coverage of the visible-to-IR wavelength range.

We derive a photometric catalog on these data by detecting objects in the F444W image and computing PSF-matched forced photometry on the remaining bands. 

We made several tests to validate the photometric calibrations, either internal, based on overlapping parts observed in different epochs with different modules, and external, based on cross-correlation with the AstroDeep catalog of the cluster region. They both confirm that photometric offset are limited to at most 0.05 mags or less. Slightly larger (0.1 mags) systematic biases, especially when $HST$ bands are concerned, could be due to the simplified PSF matching we adopt in this first release. 

We remark again that we have not explicitly removed the intra-cluster light. However, our procedure to remove the $1/f$ noise and the background on scales larger than the larger detected object are effective also in removing the ICL from the images. We therefore tested that the photometry is not significantly affected. Needless to say, this makes this dataset unfit to study the ICL and we warn interested users against using this data set for this purpose. We also note that we have not modeled and subtracted the brightest galaxy members, at variance with what we did in \cite{Merlin2016b},  so that the photometry of objects falling on their outskirt can be severely contaminated and  made brighter and generally redder.

We publicly release the entire mosaic of the NIRCam images.  The three individual images of each program, which are more homogeneous in terms of PSF orientation and coverage/depth and potentially more suitable for accurate photometry and for accurate estimate of incompleteness, are also available upon request. 

We also publicly release the  multi-wavelength catalog on the entire A2744 area, which includes 24389 objects.  We release five independent catalogs,  based on a different aperture (namely to $0.28\arcsec$, $0.56\arcsec$, $1.12\arcsec$ and $2.24\arcsec$, corresponding to  $2\times$, $3\times$, $8\times$ and $16\times$ the PSF of the F444W image) and in the isophotal area.
This catalog is optimized for high redshift galaxies, and in general for faint extragalactic sources, and aimed at allowing a first look at the data and the selection of targets for Cycle 2 proposals. In future releases we plan to include updated calibrations and procedures for the image processing and to optimize the photometry with more sophisticated approaches for PSF matching.

Finally we also release the code developed to remove the $1/f$ noise from the NIRCam images, improving the current implementation in the STScI pipeline with more effective masking of sources in the image.

Images,  catalogs and software are immediately available for download from the GLASS-ERS collaboration \footnote{\url{https://glass.astro.ucla.edu}} and \textsc{AstroDeep} website\footnote{\url{http://www.astrodeep.eu}}.
They will also be made available at the MAST archive upon acceptance of the paper. 

All the $JWST$ data used in this paper can be found in MAST: \dataset[10.17909/kw3c-n857]{https://doi.org/10.17909/kw3c-n857}.

\section*{Acknowledgement}
We warmly thank J. Weaver, K. Whitaker, I. Labb\'e and R. Bezanson for sharing their data with us prior to publication, which made it possible to compare the two processes for data analysis.
This work is based on observations made with the NASA/ESA/CSA James Webb Space Telescope, and with the NASA/ESA Hubble Space Telescope. The data were obtained from the Mikulski Archive for Space Telescopes at the Space Telescope Science Institute, which is operated by the Association of Universities for Research in Astronomy, Inc., under NASA contract NAS 5-03127 for $JWST$ and NAS 5–26555 for $HST$. These observations are associated with program JWST-ERS-1324, JWST-DDT-2756, and JWST-GO-2561, and several $HST$ programs. We acknowledge financial support from NASA through grant JWST-ERS-1324. This research is supported in part by the Australian Research Council Centre of Excellence for All Sky Astrophysics in 3 Dimensions (ASTRO 3D), through project number CE170100013. KG and TN acknowledge support from Australian Research Council Laureate Fellowship FL180100060. MB acknowledges support from the Slovenian national research agency ARRS through grant N1-0238. We acknowledge financial support through grants PRIN-MIUR 2017WSCC32 and 2020SKSTHZ. We acknowledge support from the INAF Large Grant 2022 “Extragalactic Surveys with $JWST$”  (PI Pentericci). CM acknowledges support by the VILLUM FONDEN under grant 37459. RAW acknowledges support from NASA $JWST$ Interdisciplinary Scientist grants NAG5-12460, NNX14AN10G and 80NSSC18K0200 from GSFC. The Cosmic Dawn Center (DAWN) is funded by the Danish National Research Foundation under grant DNRF140. This work has made use of data from the European Space Agency (ESA) mission 
{\it Gaia} (\url{https://www.cosmos.esa.int/gaia}), processed by the {\it Gaia}
Data Processing and Analysis Consortium (DPAC,
\url{https://www.cosmos.esa.int/web/gaia/dpac/consortium}). Funding for the DPAC
has been provided by national institutions, in particular the institutions
participating in the {\it Gaia} Multilateral Agreement. The authors thank Paola Marrese and Silvia Marinoni (Space Science Data Center, Italian Space Agency) for their contribution to the work.

\section*{Software}

\textsc{astropy} \citep{astropy2013, astropy2018, astropy2022}, 
\textsc{a-phot} \citep{Merlin2019},  
\textsc{denoise\_nircam} (\url{https://github.com/diegoparis10/denoise_NIRCam}), 
\textsc{Grizli} \citep{Brammer2019}, 
\textsc{matplotlib} \citep{hunter2007},  
\textsc{numpy} \citep{vanderwalt2011}, 
\textsc{SCAMP} \citep[][]{Bertin2006}, 
\textsc{SExtractor} \citep{Bertin1996}, 
\textsc{JWST STScI Calibration Pipeline} \citep{Bushouse_JWST_Calibration_Pipeline_V182}, 
\textsc{SWarp} \citep[][]{Bertin2002}, 
\textsc{t-phot} \citep{Merlin2015, Merlin2016a}, 
\textsc{WebbPSF} \citep{perrin2012, perrin2014}

\begin{figure*}
\center
\includegraphics[width=1\textwidth,height=0.35\textwidth]{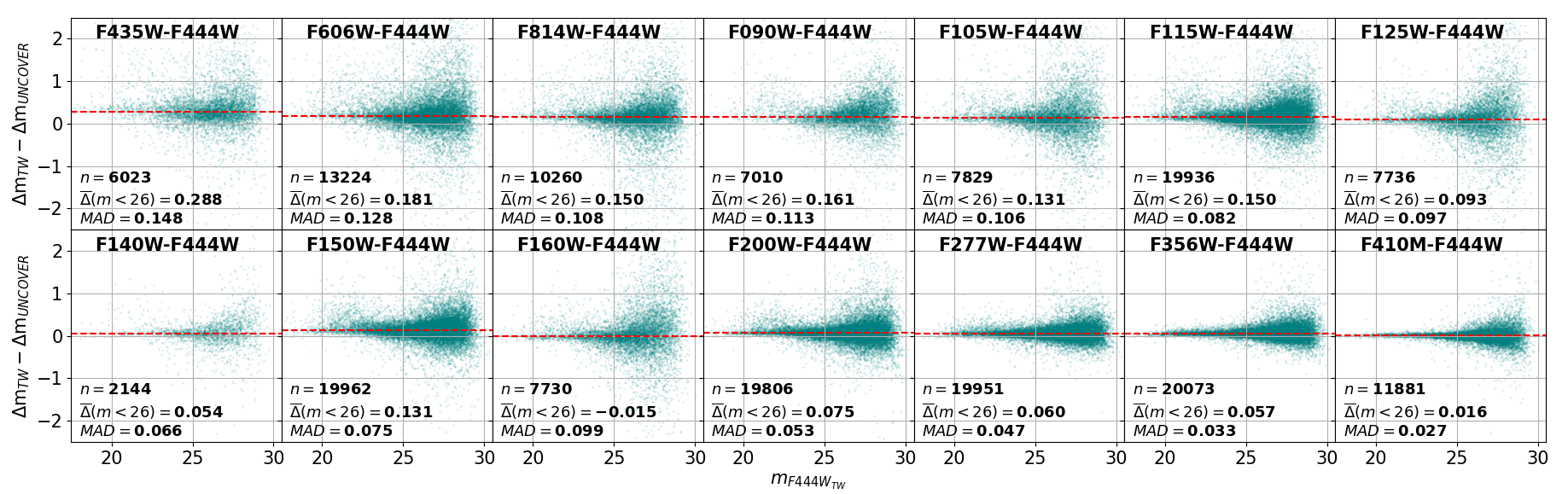}
\caption{Comparison of colors between this work (TW) and the UNCOVER catalog. $\overline{\Delta}$ represents the median offset between $\Delta m_{TW}$ and  $\Delta m_{UNCOVER}$ computed selecting only objects brighter than $m=26$ and applying a MAD-clipping to the sampled data.}
 \label{fig:catglvsunc}
\end{figure*}

\section*{Appendix}
Figure~\ref{fig:catglvsunc} shows the color comparison between our catalog and the catalog released by the UNCOVER collaboration \citep{weaver2023uncover} on the same data set, for which we take the final released version. The two catalogs have been obtained with largely independent procedures for data reduction, source detection, PSF homogenization and photometry.
In addition to the minor differences in the pipelines, for which we refer to the \citep{weaver2023uncover} paper, we explicitly remark that their procedure has explicitly removed ICL and brightest cluster members, and adopted different recipes for PSF estimation.
Despite these differences, we find a general good agreement between the two catalogs, in particular for the $JWST$ long wavelength but also for most of the other bands. There seems to be a systematic trends of our catalog yielding redder colors as we move to shorter bands. We explicitly notes that disagreement of the same amount - often even larger - are found when comparing catalogs extracted from previous HST surveys, despite the fact that processing of HST data is certainly more established and accurate than the JWST one, due to lack of complete calibration. While it is tempting to associate these deviations to different PSF matching techniques, we plan to investigate them in more details in our final analysis of this data.

\bibliography{new.ms.bib}
\bibliographystyle{aasjournal}

\end{document}